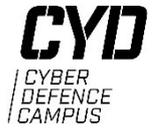

26 March 2025, Cyber-Defence Campus

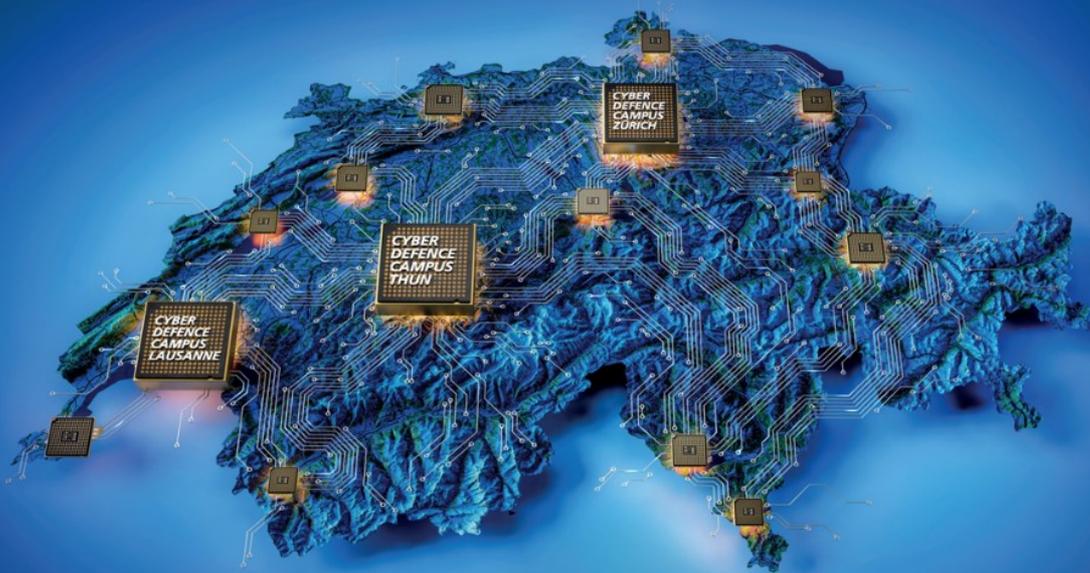

# Threats and Opportunities in AI-generated Images for Armed Forces

Raphael Meier

## Management Summary

Images of war are almost as old as war itself. From cave paintings to photographs of mobile devices on social media, humans always had the urge to capture particularly important events during a war. Images provide visual evidence. For armed forces, they may serve as the output of a sensor (e.g. in aerial reconnaissance) or as an effector on cognition (e.g. in form of photographic propaganda). They can inform, influence, or even manipulate a target audience. The recent advancements in the field of generative Artificial Intelligence (AI) to synthesize photorealistic images give rise to several new challenges for armed forces.

The objective of this report is to investigate the role of AI-generated images for armed forces and provide an overview on opportunities and threats. We start this journey with a discussion of a series of technological breakthroughs, including photography, mass media, social media, and generative AI, that contributed to the vulnerability of today's information environment. We explore the added value of AI-generated images for the two most basic military functions when it comes to images: their use as sensor output, and their use as effector on cognition. Based on this, the conceptual advantages of AI-generated (synthetic) images will be contrasted with real images generated by traditional means (e.g. photography). Here, generative AI offers text prompts as an easy-to-use mechanism to condition the generation process and thus providing actors with great flexibility in tailoring their visual content to a particular message or concept. They also remove the need for actors to repurpose old image material, thus increasing their operational security. When compared with traditional image generation, candidate images can be generated more efficiently (no need for photographic equipment, setup of a scene, skilled photographer, dependence on illumination, etc.). Furthermore, the programmatic nature of generative AI algorithms provides a basis to upscale the generation of images to very large quantities of images. These conceptual advantages combined offer the opportunity to implement new tactical tenets and concepts which so far have not been feasible: masses of AI-generated images can be used for deceptive purposes, to influence the pace of combat in the information environment, to cause surprise, sow confusion and shock. AI-generated images are a tool favoured for offensive manoeuvres in the information environment. This is also reflected in our review of opportunities and threats which identifies several areas of deception, manipulation, and subversion that are greatly affected by the introduction of AI-generated images including cyber influence operations, social engineering, and military deception. The report is concluded with a set of recommendations for all branches of the armed forces, who are active in cyber defense and/or exposed to the information environment, on how to best prepare for future challenges involving AI-generated images.

# Table of Contents

# 1 Introduction

Images and warfare have a long, intertwined history. We find depictions of war ranging from neolithic cave paintings to today's online propaganda on social media (see Figure 1). Images are visual representations. They can be drawn, painted, sculpted, carved, or photographed. The advent of photography and advancements in camera technology facilitated the accurate depiction of real objects and events (e.g. the enemy's force disposition photographed from the air). This capability was recognized early on for its great military potential, and it was thus investigated in a series of experiments within armed forces beginning already in the mid-19$^{th}$ century. Early applications included the documentation of armed conflict through war photography and aerial reconnaissance using ballons. Later in World War I, two basic military functions of images developed: their role as sensor output, and their role as effectors on cognition (e.g. as a medium for propaganda). Images, particularly photographs, could in part hold information relevant for military intelligence but also propagate messages to influence target audiences. Information has long been regarded as an instrument of power for nation states and as such is integrated in a variety of different military doctrines (e.g. NATO's AJP-01, *Allied Joint Doctrine*[1]*)*. Information plays a major role not only during armed conflict but particularly also for hostile actions during confrontations below the threshold for armed conflict used in cyberspace and the information environment. When we think of the information environment as "*an environment comprised of the information itself, the individuals, organizations and systems that receive, process and convey the information, and the cognitive, virtual and physical space in which this occurs*"[2], images are a tool to capture and represent part of the information that forms the information environment. In this context, photographs hold a special place. Photographs have for a long time been regarded as a medium that can hold authentic information. Early on photographs were compared with drawings and paintings, and were being regarded as "unmediated" [1], promising an authentic depiction of the real. However, advancements in camera technology and photography led also to new ways of manipulating photographs. As such they can both be a source of credible intelligence and of subtle manipulation. Recent advances in generative Artificial Intelligence (AI), which enable the generation of whole photorealistic synthetic images, are challenging the authenticity of photographs now more than ever. Consequently, generative AI offers new threats and opportunities for armed forces in both offensive and defensive manoeuvres.

The objective of this report is to provide the reader with an overview on the threats and opportunities for armed forces arising from the increased availability of AI-generated images, particularly photorealistic images. In order to equip the reader with the necessary background information, the report will first give a brief overview on the military significance of images, their basic military functions and how generative AI is currently transforming them. In doing so, we will focus on the two basic military functions outlined previously: images as sensor output, and images as effectors on cognition. The later includes also the use of images to support education and training of troops. Furthermore, the key conceptual differences of AI-generated (i.e. synthetic) images and real images will be investigated including their implications for armed forces. Based on insights from research literature and actual case studies, several threats and opportunities for armed forces will be reviewed. This report is seen as an introduction into the topic based on a selection of the most relevant and recent scientific and military literature. It is by no means a systematic literature review and may thus omit additional references for the sake of brevity. Readers interested in systematic literature reviews of adjacent topics are referred to the respective published surveys (e.g. on AI methods in text-to-image generation [2], on ethics of generative AI [3], or on safety of Advanced AI [4]).

---

[1] Allied Joint Doctrine (AJP-01) - GOV.UK [accessed December 2024]
[2] Quote from Allied Joint Doctrine for Strategic Communications (AJP-10) - GOV.UK [accessed December 2024]

## 2  Military Significance of Images

In the following, the military significance of images is being introduced through a presentation of a series of technological breakthroughs that had a lasting effect on the conduct of war over time and the role images played in it. This journey will lead to several observations that highlight the vulnerability of today's information environment. The role of images as sensor output and as effector on cognition will be introduced, which reflect their most basic military functions. This chapter provides the basis to introduce the conceptual advantages of synthetic versus real images for the armed forces in the subsequent chapter.

### 2.1  History & Technological Breakthroughs

When we look at the history of war, we find visual depictions thereof, images, from the very beginning of human history until today. War is a gruesome endeavor, often traumatizing to the people involved in it. According to Clausewitz 'war is a mere continuation of policy by other means'. This emotionally laden political function may be the driving force behind image creation. From psychology, we know that trauma can be overcome through creation of art (in art therapy [5]). In addition, what we also know from psychology is that images have diverse cognitive effects on their audience, they have the ability to *manipulate* [6]. Whatever the motive, be it curative (for the creator) or manipulative (to the audience), an image provides *visual evidence* that may complement oral or written narration. This can be regarded as the core functionality of images in context of war. It is common to find images of particularly significant events in a war. For example, images that show the execution of traitors can be found throughout human history (see Figure 1). Images may also facilitate comprehension of complex issues (e.g. in education of new recruits) and can serve as a representation of factual information (e.g. in case of military maps). The first two images in Figure 1 are paintings, whereas the third image is a photography. The acquisition, distribution and consumption of images has since the very beginning be subject to a continuous transformation driven by a series of technological breakthroughs. In this chapter we will review the most important technological breakthroughs and the impact they had on images in context of war.

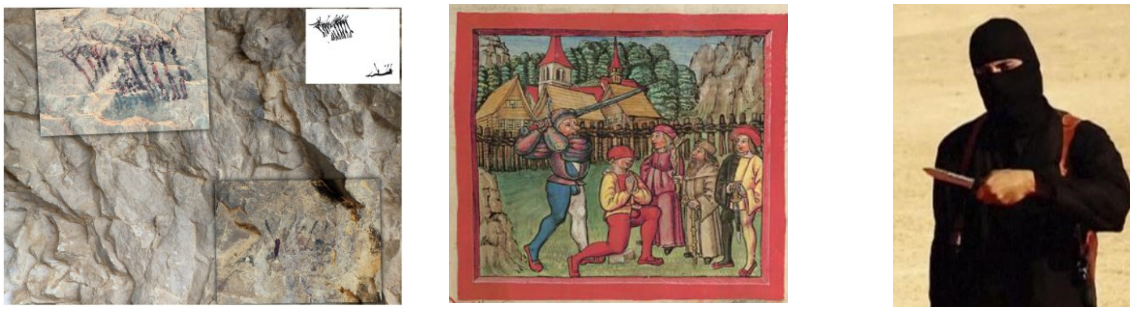

Figure 1.  Images that provide evidence on the execution of traitors during times of war can be found throughout human history from neolithic cave paintings (Cova Remigia-V, Gasulla, Castellón, Spain, figure on the left adapted from [7]), to medieval illustrations (execution of Hans Turmann, Luzerner Chronik, 1513, figure in the middle adapted from https://hls-dhs-dss.ch/articles/008894/2010-09-09/ [accessed November 2024]), to propaganda material of the Islamic State on social media (figure on the right adapted from wikipedia.org).

The first major technological breakthrough took place in the mid-19$^{th}$ century with the introduction of photography. In Figure 2 we see two early photographs acquired in 1855 during the Crimean War between the British and Russian empires. The photographs were taken by Roger Fanton[3], a wealthy British lawyer, who travelled to Crimea with his photographic van in the fashion of a very early version of an embedded war photographer. At that time photography was contrasted with paintings and its significance was being debated. In general, paintings were seen as an idealization of reality. Photographs were seen as "unmediated", they functioned as "witnesses" of history [1]. This perception stays in stark contrast to the complicated handling and very long exposure times of photography at that time which necessitated that photographed persons remain still for a relatively long period of time.

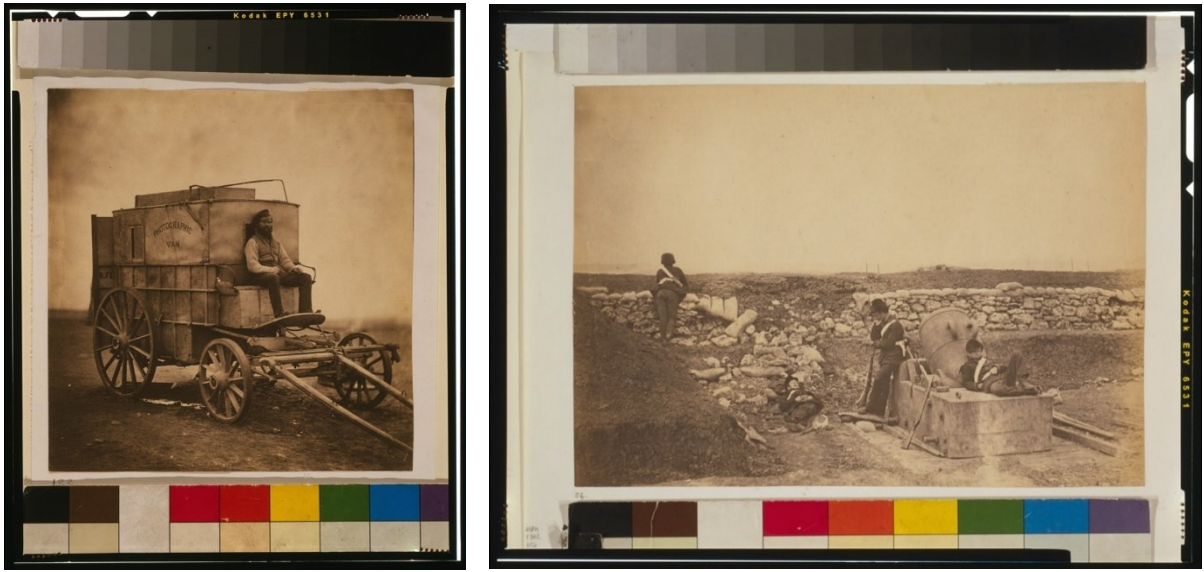

Figure 2. Photographic van of Roger Fanton in 1855 during the Crimean war (left image, The Roger Fenton Crimean War Photograph Collection, Library of Congress, Prints & Photographs Division, LC-USZC4-9240). Early photography by Roger Fanton of a British mortar battery near Sevastopol during the Crimean war (right image, The Roger Fenton Crimean War Photograph Collection, Library of Congress, Prints & Photographs Division, LC-USZC4-9170).

Over the years photographic technology developed further and eventually during World War I became integrated into the first mature military functions. In particular, the combination of photography with airborne platforms like balloons or planes led to an increased use of aerial reconnaissance throughout the war (see Figure 3). The analysis of images from aerial reconnaissance missions was slowly being institutionalized in military intelligence with courses on aerial reconnaissance and photography being offered by e.g. the British Intelligence School in Harrow and the American Expeditionary Forces (AEF) Intelligence School in Langres at the end of the war [8]. Aerial reconnaissance provided crucial topographic information that could be combined with geographic information from military maps to support both defensive and offensive operations on the ground. Photography found its role as part of a sensor.

---

[3] . [Photographing conflict: Roger Fenton and the Crimean War - National Science and Media Museum blog](#) [accessed November 2024]

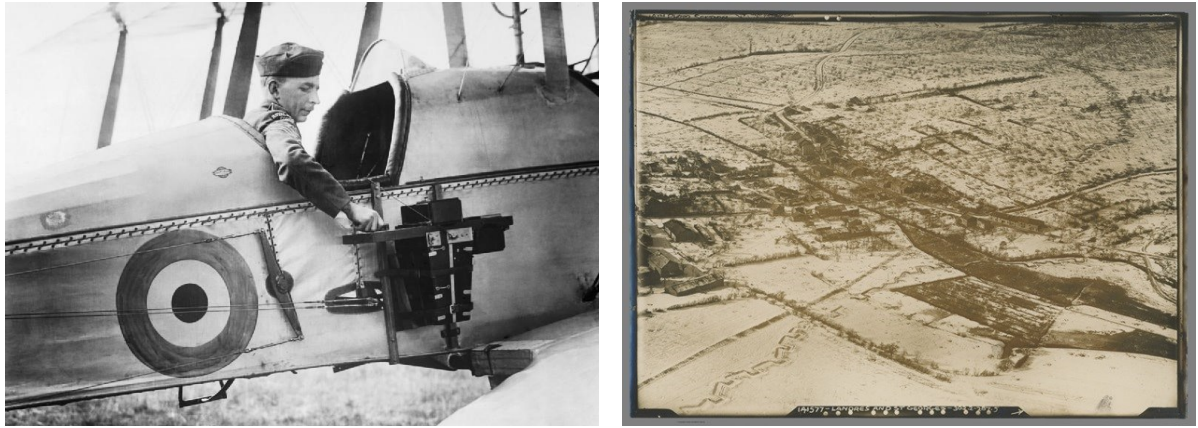

Figure 3. Camera mounted on a British B.E.2c reconnaissance airplane (left image, image taken from wikipedia.org). Aerial photograph from an oblique angle of a small village in the Ardennes during World War I by an Aerial balloon of the American Expeditionary Forces (AEF) (right image, image credit: David Rumsey Map Collection, David Rumsey Map Center, Stanford Libraries). Topographic information extracted from aerial photographs was useful when combined with traditional cartographic information from military maps. Airborne cameras evolved into a sensor to collect information for military intelligence.

It was also during World War I that photography was recognized as a tool for propaganda[4]. Photographs cause various cognitive effects on a target audience, as such they can be regarded as an effector. The use of photographs as effector on cognition required the reproduction of the same image in very large quantities to facilitate distribution among the target audience. This became possible in the late 19$^{th}$ century with the introduction of rotary offset lithography, which was patented by Robert Barclay in 1875. Photographs were printed in newspapers and could thus potentially reach a large target audience. In Figure 4, we see two early examples of photographic propaganda taken during World War I. The first example was taken by John Warwick Brooke, a British official war photographer who was sent to the Western Front. It shows British soldiers who purportedly occupy a captured German trench. The rather relaxed appearance of the soldiers may be an indication that the photograph was staged, which was the main tactic to condition the image generation process. The second example shows an early manipulated photograph acquired by the Australian Frank Hurley during the Battle of Passchendale, which exhibits a doctored sky, highlighting the fact that image manipulation was already possible at that time. The ability to influence the content of the final photograph, either through staging or editing it, was crucial to exchange authentic information with information containing a hidden agenda—a prime feature of propaganda[5]. In addition to propagandistic uses, photographs were also slowly being employed by armed forces to support education and training activities of troops, e.g. in weapons manuals, combat first aid training, etc.[6]

---

[4] An Image is Worth a Thousand Words | American Battle Monuments Commission [accessed September 2024]
[5] The Yale Review | Renée DiResta: "Computational Propaganda" [accessed February 2025]
[6] Field Manuals – All Americans [accessed January 2025]

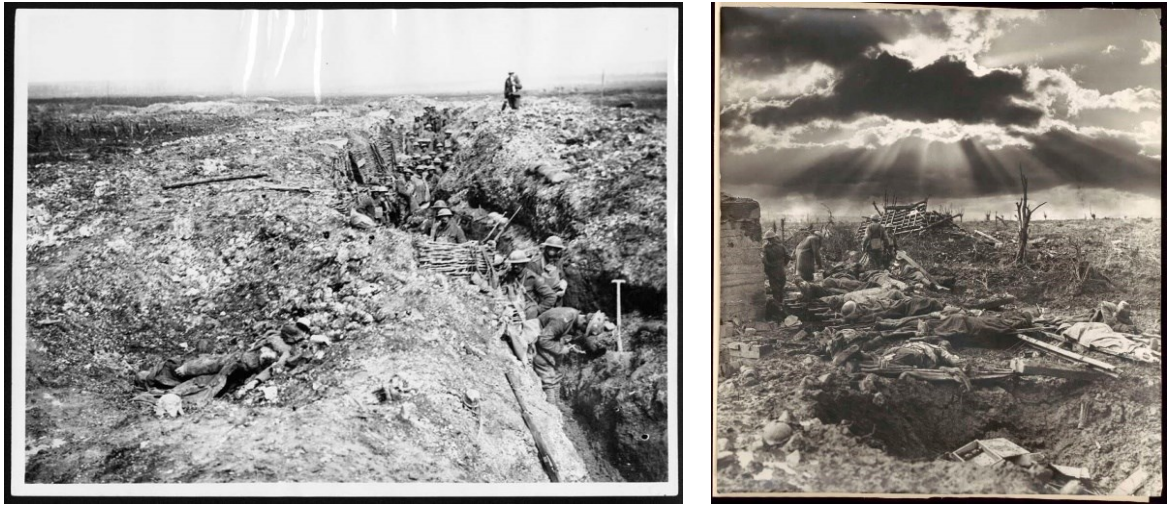

Figure 4.  British soldiers occupying a German trench during World War I in a potentially staged image taken by the official British war photographer John Warwick Brooke (left image, image taken from https://digital.nls.uk/first-world-war-official-photographs/archive/74547526 [accessed February 2025]). Picture of Australian soldiers during the Battle of Passchendale showing signs of early image manipulation (doctored sky, image taken from https://nla.gov.au/nla.obj-147383187 [accessed February 2025]).

The next major technological breakthrough did not affect the acquisition of images, like photography did, but their distribution: mass media. Newspapers, radio and television (TV) enabled the rapid dissemination of information to a large target audience. Information also flowed across channels from newspapers to radio and TV broadcasts. Mass media outlets frequently referenced each other. This led to the development of a new tactic during the cold war era, called "information laundering". It is best explained by the famous example of a soviet influence operation, called Operation Infektion[7]. The aim of the operation was to spread a conspiracy theory that AIDS was developed in a biological laboratory of the US army to selectively target the black and homosexual populations worldwide. This conspiracy was initially planted in an Indian newspaper in 1983, from where it spread to press media of other countries in Africa and South America. The conspiracy was further supported by a hoax publication of two East German biologists in 1986 who claimed to have proven that AIDS originated in a laboratory (see Figure 5). In 1987 it was referenced in a CBS broadcast potentially reaching a large proportion of the American population. This example highlights how a message can travel across different channels such that it becomes hard to attribute it to its original source. Furthermore, the message may gain credibility and it increases its reach as it spreads from unverified sources to mainstream media outlets.

---

[7] Operation Denver - Wikipedia [accessed February 2025]

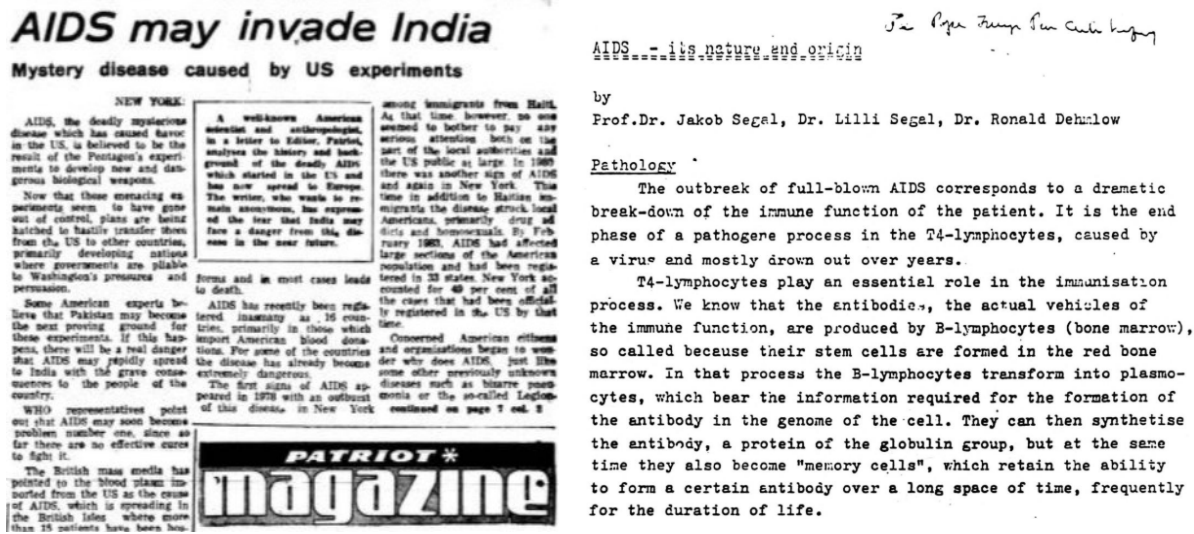

Figure 5. Initial article of the Soviet influence operation "Operation Infektion" which spread the conspiracy that AIDS was developed in a biological laboratory of the US army (left image, image taken from https://archive.org/details/1983-07-16-patriot/mode/2up [accessed February 2025]). Hoax publication of two East German scientist who claimed to have proof of AIDS originating in a laboratory (right, image taken from https://archive.org/details/AIDS-nature-and-origin/mode/2up [accessed February 2025]). As a message travels across different channels and platforms, it may gain credibility and hide the original source of the information (information laundering).

The advent of social media constituted the next technological breakthrough. Social media facilitated a democratization in dissemination of information in the late 2000s and early 2010[th] which was unprecedented. In combination with the increased availability of smartphones, it enabled sharing of image and video material among arbitrarily large groups of people. Traditional gatekeeping mechanisms, which are present in mass media, are absent. This led to more variety in the topics being discussed online but also to an abundance of low-quality information [9]. Furthermore, the shared media data could easily be discussed with other people and forwarded to people unaware of it, thus fostering online discourse. Social media increased the reach of media content across national borders, different communities and cultures[8]. It greatly facilitated collective sensemaking processes for social groups during times of crisis (e.g. Arab Spring in 2011 [10]). In 2014 the Islamic state recognized social media as an opportunity to conduct an influence operation during their march to Mosul. They deliberately amplified image and video content showing convoys of trucks transporting armed men and their brutal course of action over a single hashtag #AllEyesOnISIS [11]. It is hypothesized that this influence operation contributed to the low morale of the Iraqi defenders of Mosul. Furthermore, social media only increases the opportunities for information laundering through the availability of multiple platforms, the ease with which a user account can be created, and the ability to cross-reference media content among platforms (e.g. sharing a Tweet from X in a Telegram channel, which is commented, screengrabed and subsequently shared on Instagram). As a message spreads from one platform to the other, it may gain credibility. This phenomenon has been formalized by the breakout scale [12], which measures the impact of a social media influence operations' messages through the number of platforms and communities they spread to (from affecting one community on one platform to multiple

---

[8] Hybrid CoE Strategic Analysis 5: Beyond Fake News: Content confusion and understanding the dynamics of the contemporary media environment - Hybrid CoE - The European Centre of Excellence for Countering Hybrid Threats [accessed November 2024]

communities on multiple platforms, and eventually to mainstream media and high-profile individuals).

The most recent technological breakthrough is generative Artificial Intelligence (AI) [13]. It includes a subset of methods from the field of AI which are trained on large datasets to generate text, image, audio, and/or video data. These methods produce models that learnt the underlying distribution of the training data and use this information to generate new samples. The generation of synthetic image data was taken to a new level of quality with the introduction of diffusion models [14]. This approach is based on two basic processes called forward and reverse diffusion. In forward diffusion random noise is iteratively added to an image. During model training the model is tasked to subtract random noise in successive steps to regenerate the original image, which is called reverse diffusion. The resulting model is able to generate synthetic images from a random noise vector. This approach was soon enhanced with text encoder which provided it with a mechanism to condition the image generation on text data (see Figure 6). It turned out to be a crucial innovation since text offers an easy-to-use interface, accessible to a large proportion of the population, to personalize and guide the image generation process. As a result, the generation of sophisticated visual content has been increasingly democratized.

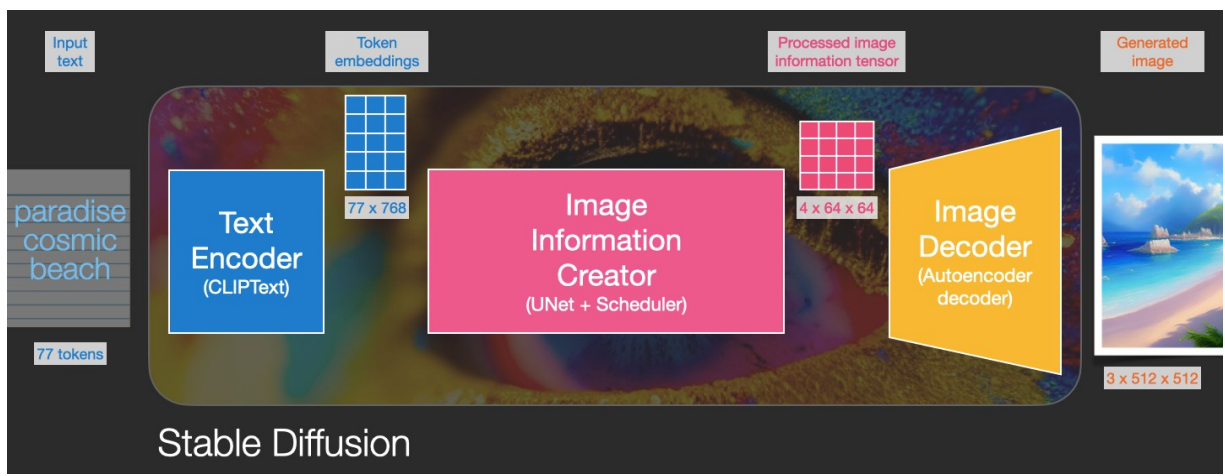

Figure 6. Basic components of a text-to-image diffusion model (Stable Diffusion) including text encoder as the main interface to condition the image generation process based on user-defined text prompts (image taken from https://jalammar.github.io/illustrated-stable-diffusion/ [accessed November 2024]). The integration of text encoders offers an easy-to-use interface for a large proportion of the population. Thus, it democratizes the generation of visual content.

We can draw several observations from the previously discussed history and breakthroughs:

- Photographs may function as either part of a sensor or effector. They provide visual evidence and may cause different cognitive effects (e.g. to attract attention [15]).
- Social media led to an unprecedented democratization of dissemination of information. Today, we have large quantities of low-quality information.
- The combined infrastructure of mass media and social media facilitates information laundering. Source attribution is becoming harder and harder.
- Generative AI caused a democratization in creation of sophisticated visual content. For publicly-available, open-source models, there is no gatekeeper, no filters and no restrictions on what type of content can be generated.

All these observations combined imply that we live in a time with an extraordinarily vulnerable information environment. This is particularly relevant for open societies in democratically ruled countries, for which the vulnerability of the information environment is systemic. In the following, we want to briefly outline the nature of this vulnerability and its implications for democratic societies and their armed forces.

## 2.2   Vulnerable Information Environment

Hybrid threats include the unacceptable foreign interference by a threat actor in a sovereign state's internal affairs and space[9]. When conducted against democratic societies, they usually target their systemic vulnerabilities. Hybrid threats rely on covert actions below the threshold of conventional warfare and typically combine different tools (e.g. espionage to collect sensitive information which then can be purposefully disseminated in a cyber influence operation to manipulate public opinion). Results of hybrid threats may include influenced decision-making of the target state's systems towards a desired outcome (e.g. manipulated election, new policies or legislation, business deals, etc.). Information can serve as an entry point for different attack vectors in hybrid threats[10]. An adversary can use selected information to manipulate, deceive, or subvert target individuals, groups, or institutions. Well-informed citizens are regarded a prerequisite for functioning democratic societies [16] and are therefore a prime target for information manipulation. Figure 7 shows a prototypical attack vector to influence a target audience and/or media agenda. First, the adversary creates and cultivates a synthetic persona on one or multiple social media platforms. Second, generative AI is used to create media content that fits a particular topic (e.g. immigration) which should be messaged with a clear objective in mind (e.g. increase polarization in online discourses, elicit fear or anger in target audience). The message is then distributed across different social media channels, where it is eventually picked up and amplified by authentic users. This is particularly effective when amplification is performed by an influencer and/or large organic online crowds [17] [18]. Consequently, it reaches a large part of the target audience and may also influence the agenda of mass media to the advantage of the adversary. This can be done quasi anonymous, is cheap and potentially scalable. Potential roadblocks for threat actors are the access to social media platforms and creation of credible synthetic persona with sufficient reach. Such activities are typically performed covertly and can be summarized with the term cyber-enabled social media influence operations (CeSIOs) [19]. There are a multitude of frameworks available which further characterize and conceptualize this threat (e.g. the breakout scale mentioned earlier). More broadly, we can speak about cyber influence activities which also include activities targeted at individuals (e.g. social engineering).

---

[9] Definition taken from The landscape of Hybrid Threats: A conceptual model - Hybrid CoE - The European Centre of Excellence for Countering Hybrid Threats [accessed November 2024].
[10] Based on the framework introduced in JRC Publications Repository - Hybrid Threats: A Comprehensive Resilience Ecosystem [accessed November 2024].

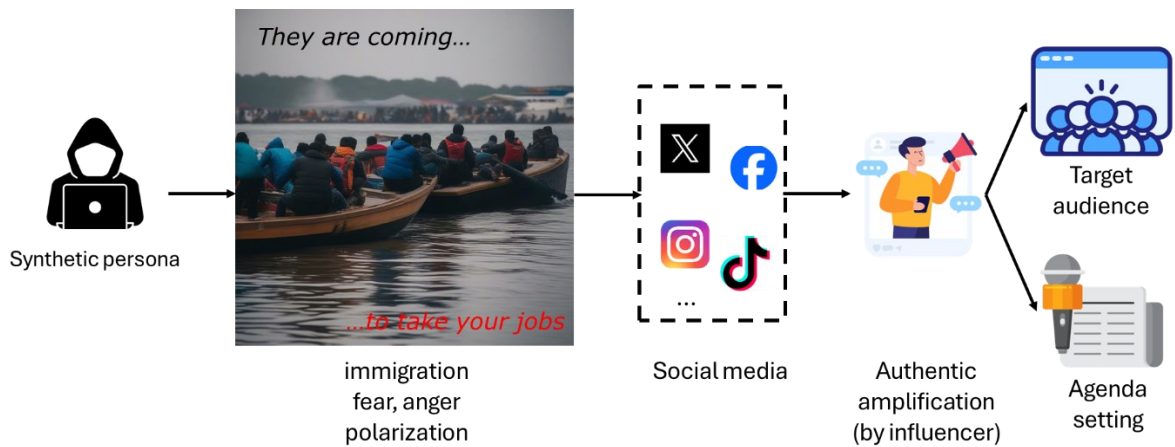

Figure 7. Prototypical attack vector in a cyber influence operation (photograph generated with Stable Diffusion XL, icons from Flaticon.com, made by Talha Dogar, lutfix, Flat Icons, smashingstocks, and from wikipedia.org). The threat actor creates and cultivates a synthetic persona through sock puppet accounts on different social media platforms. The sock puppet accounts will be leveraged to propagate a particular message on a chosen topic (e.g. immigration) with a set of predefined criteria (e.g. spread fear and anger on the topic, to polarize discourse). Generative AI can be used to support the message with visual content. When the message is amplified via a crowd of authentic users and/or influencer(s), it reaches the target audience and may also influence agenda setting.

In the following, the focus will be on the implications of such a vulnerable information environment for armed forces. An adversary can utilize the attack vector presented in Figure 7 to influence the public perception of the armed forces or to weaken their moral cohesion. For example, the Polish Armed Forces were targeted by Russian cyber influence operations during the COVID-19 pandemic with the supposed aims to undermine the morale of the Polish troops, to present the armed forces as a threat to their own citizens, and to hurt relations with the US military deployed in Poland[11]. Since the start of the full-scale invasion of Ukraine, Ukraine's recruitment efforts have been the target of continuous Russian cyber influence operations to undermine mobilisation of new troops.[12] During the Hawaii wildfires in August 2023, actors aligned with the Chinese Communist Party (CCP) propagated AI-generated images of burning coastal roads and residential areas aimed at fueling a conspiracy theory that the fires were caused due to a test of a secret "weather weapon" by the US government[13]. All these examples have in common that they rely on synthetic persona (or fake news outlets) propagating fabricated, partially AI-generated media content over multiple social media channels with the aim to create some authentic amplification of their messages as outlined in Figure 7. The same attack vector may be used to "increase the entropy" for the opponent's armed forces through polluting the information environment with false or irrelevant information. This is particularly significant for troops who incorporate the principle of the OODA-Loop. The first step of the OODA-Loop is "observe" and is tasked with the collection of relevant information. Today, open-source intelligence (OSINT) and commercially-available information (CAI) play an important role in information collection for armed forces. Large quantities of false or misleading information can hamper an armed forces' capability to perform OSINT or extract intelligence

---

[11] Propaganda attacks on the Polish Armed Forces - Special Services - Gov.pl website [accessed November 2024]
[12] How Russian special information operations try to undermine mobilisation in Ukraine [accessed November 2024]
[13] Same targets, new playbooks: East Asia threat actors employ unique methods | Security Insider [accessed November 2024]

from CAI. This may lead to a slower OODA-Loop and potentially to military defeat. ISIS' influence operation during their march to Mosul can be seen as an example of this. The mass of the posted image and video footage conveyed a situation with inflated number of troops and vehicles approaching from multiple directions, which probably helped to paralyze the Iraqi armed forces' decision-making processes.

In order to round out the significance of images for armed forces, we want to briefly discuss the two elementary functions of images and discuss the added value of AI-generated images for each of them.

## 2.3  Images as Sensor Output

As shown in the previous example on aerial reconnaissance in Figure 3, images can be part of a sensor. In particular, images often form the *output* of a sensor (e.g. images taken with a camera device). As such images hold information that can be of critical importance for the armed forces (e.g. concentration of enemy troops or novel construction of defensive fortifications visible in satellite imagery). Raw image information is collected, processed, and analyzed with the objective to provide intelligence for military decision-making processes. Images as sensor output may also be the input to an effector (e.g. camera-guided first-person view drone).

Generative AI will impact machine learning (ML) models tasked with the analysis of sensor output images (e.g. object detection algorithms). The training data of these ML models may be augmented by synthetic images generated with the help of generative AI. Several research studies have shown improvements of ML models when trained on synthetic data in a diverse range of applications (e.g. vessel classification using thermal infrared images [20], aircraft detection from satellite imagery [21], etc.). Therefore, armed forces can use synthetic, AI-generated image data to improve their set of ML models tasked with solving different problems from computer vision. Compared with the next function (images as effectors), the impact of generative AI for this function will probably be constrained to the domain of synthetic training data. However, it is important to note that the use of synthetic training data is not without any risks. It has been shown that synthetic data can also deteriorate the performance of ML models [22] and must therefore be carefully integrated in the ML model's training procedure. For operatives of armed forces confronted with the question of whether to use synthetic training data or not, this implies that a high competency in training of ML models, machine learning theory, and data analysis is necessary, to answer this question and to use this data to yield actual improvements in a given application.

## 2.4  Images as Effectors on Cognition

As outlined previously, images have cognitive effects on their audience. This can be exploited by adversaries e.g. in form of visual disinformation propagated in a cyber influence operation. However, images as effectors also serve a role in education and training of the armed forces personnel. Insights from psychology showed that images, particularly with moral/emotional content [23], draw attention, are more likely to be shared and re-shared than text alone [15], and increase the perceived accuracy [24] of a message (irrespective of its source [25]). Moreover, it has been shown that the news frame (i.e. central organizing idea of a news article) carried by an image generates a stronger effect than text alone, and for image-text pairs it drives the audience's behavioral intentions irrespective of the linked text (e.g. the intention to discuss, donate to a cause, sign a petition, or participate in a protest) [26].

The impact of generative AI will be likely substantial for this function. Generative AI will serve as a force multiplier for any activity involving the distribution of image content for the purposes of deception, manipulation and subversion. In fact, generative AI features distinct advantages compared with traditional image generation methods as detailed in the next section. It offers great flexibility in creation of novel synthetic image content through various mechanisms of conditioning the generation process. Therefore, threat actors can tailor content with a high degree of flexibility and without the need to repurpose existing visual material. This will make it significantly harder to (soft) attribute image content to actors (e.g. through reverse image search) increasing their operational security. Furthermore, it has been shown repeatedly that humans struggle to recognize synthetic images and differentiate them from real images. In a study by Lu et al. [27], participants showed a misclassification rate of 38.7% for synthetic versus real photographs. Di Cooke et al. [28] observed a ~50% misclassification rate for 96 synthetic images (including photographs of human faces) in 1276 participants. Ha et al. [29] found that general (non-artist) users are unable to distinguish real from AI-generated visual art (40.77% misclassification rate). Older persons appeared particularly vulnerable with increased misclassification rates (e.g. in discrimination of synthetic vs. real photographs of unprocessed/processed food [30]). Due to its algorithmic nature, generative AI can be scaled up to generate large quantities of synthetic images and/or generate different variations of a visual message repeatedly. Combined with the insights from psychology outlined in the beginning of this section, the process of image generation can thus be repeated many times resulting in the exposure of a target audience to the same message repeatedly, profiting from increased attention and audience reach through visual stimuli and increasing the likelihood that the message is being believed ("illusory truth effect") [31]. As such AI-generated images can be seen as a force multiplier for cognitive effects mediated by the information environment. A more detailed discussion on the advantages and limitations of AI-generated images for threat actors in cyber influence operation will be given in Section 4.5.

## 3    Conceptual Advantages of Synthetic vs. Real Images

In Section 2 we saw that images have been used in context of warfare for a very long time. Therefore, it is important to investigate the conceptual advantages of AI-generated (or synthetic) images when compared with real images. The starting point for this investigation is formed by what we call *image generation task*. The objective of this task is to generate an image that fits a particular purpose (e.g. portrait photograph of a specific individual) and/or a set of predefined criteria (e.g. realism, adherence to a topic, emotionality, etc.). The task can be solved through either the use of generative AI, yielding synthetic images, or the use of "traditional" image generation (e.g. photography and image editing). When comparing generative AI with traditional image generation, it offers two distinct advantages: i) a relatively simple to use mechanism for conditioning of the image output, and ii) scalability. In the following, we will detail these advantages and introduce possible tactical tenets and concepts that can be implemented through the use of generative AI.

### 3.1    Mechanism for Conditioning

Generative AI offers a number of different interfaces to its users which enable the conditioning of the image generation process based on other, usually image or text, data. In particular, text-to-image models offer textual prompts as an intuitive and easy-to-use interface to impose different conditions on the generation process. Text-to-image models map textual prompts to synthetic images through the means of an AI-driven generation process (e.g. diffusion). This process is influenced by several user-defined parameters which control the outcome (e.g. cfg guidance scale, which controls the adherence of Stable Diffusion, a publicly-available image-to-text model, to the textual prompt). We can regard a single AI-generated image as a single point in a large space of potentially valid and useful generations (depending on the chosen parameter setting). Furthermore, every change made to the textual prompt will yield a different AI-generated image and as such a different point in this space. To find the most optimal image, users of generative AI typically engage in what is known as "prompt engineering", i.e. the careful adaptation of the prompt to generate a candidate image that fits a set of predefined criteria (e.g. the realistic depiction of a public figure in a compromising context as a malicious use case). Hence, we can regard the space of potential AI-generated images as the solution space for the image generation task.

When we compare the iterative AI-driven generation of images and subsequent refinement of prompts with traditional image generation, e.g. photography and image editing, it becomes evident that the solution space for AI-generated images can be *accessed more quickly* (definition of a prompt, trigger of generation) than in traditional image generation (setup of photographic equipment, setup of scene, taking photograph, editing). In other words, generative AI can produce a solution more quickly than traditional image generation. It also offers greater *flexibility* in content creation without the constraints of traditional image generation (e.g. dependence on illumination, equipment, etc.). Thus, generative AI replaced the cumbersome physical aspects of image generation with a computational generation process run on dedicated computer hardware. The relatively quick generation of different AI-generated image candidates may help in the exploration of the solution space. Depending on the available computer hardware, the generation of a synthetic image using Stable Diffusion is a matter of seconds. Thus, it may help in reduction of uncertainty in any task in which the objective depends on combined analysis of individually generated images (e.g. synthetic training data for object detection algorithms). However, it may also increase uncertainty in any task which is based on fidelity of image information (e.g. military deception through spread of synthetic satellite images).

The faster image generation process of generative AI comes also at a price. As previously mentioned, the generation process is usually parametrized by a number of different parameters (e.g. algorithmic parameters, model checkpoints, etc.) whose impact on the generated image may be difficult to interpret. In contrast to traditional image generation, the output of good quality images that fit the intended purpose or predefined criteria is by no means guaranteed. The generation of synthetic images is usually a process marked by trial-and-error. While the entry for new users to operate text-to-image models is relatively easy, particularly facilitated through dedicated Graphical User Interfaces (e.g. Automatic1111[14]) and an abundant number of online tutorials, effective use of generative AI is hard to master and requires considerable experience to generate realistic-looking images. In particular, convincing photorealism beyond mere portrait photographs is still hard to achieve. AI-generated photographs feature a broad variety of visual artifacts that can be grouped in six different categories: physical artifacts, geometrical artifacts, artifacts in human anatomy, distortions, artifacts in text, and semantic artifacts [32]. The correction of visual artifacts after generation of the images involves image editing techniques (e.g. inpainting) and often results in the introduction of new artifacts.

## 3.2   Scalability

A byproduct of generative AI's mechanisms for conditioning is their potential for scalability. Due to their algorithmic implementation, it is in principle possible to programmatically upscale the generation of candidate images. When the intended purpose or the predefined criteria of the image generation process can be implemented algorithmically and measured in a quantitative fashion, the AI-generation of images can be scaled up easily. For example, if the intended purpose is to generate a synthetic photograph showing a person, the presence of a person could be verified automatically through the means of an object detection algorithm. However, the assessment of realism, fit for purpose and overall image quality is a non-trivial, still open research problem. Hence, the generation of large quantities of realistic-looking good quality synthetic images is still very challenging. Nevertheless, with further improvements in this line of research, we can expect to see actors adopting tactics that build on *large masses* of synthetic images. This brings us to a discussion on the tactical tenets that can be implemented through the availability of AI-generated images.

## 3.3   Tactical Tenets & Concepts of Synthetic Images

AI-generated images can be regarded as a tool that can be leveraged by an actor for defensive and offensive maneuvers in the information environment. Based on the conceptual advantages of generative AI over traditional image generation processes, which were explained previously, we will now characterize the impact AI-generated images may have on the tactical level. The nature of generative AI makes it suitable to implement a set of basic tactical tenets. Based on the tactical tenets, concrete tactical concepts for AI-generated images can be hypothesized. For doing so, we build heavily on the ideas introduced by B.A. Friedman in his seminal book "On Tactics" [33]. In contrast to Friedman's proposal, we do not differentiate between physical, mental and moral tenets but rather try to translate the respective tactical tenets to an equivalent in the information environment.

As already hinted in the previous subsection, the generation of synthetic images can in principle be automated. The only roadblock is the assessment of quality and fit for purpose

---

[14] [GitHub - AUTOMATIC1111/stable-diffusion-webui: Stable Diffusion web UI](#) [accessed September 2024]

which is currently still performed by a human operator. If we assume that this step can be sufficiently automated, arbitrarily large masses of AI-generated images could be generated and deployed. Hence, the tactical tenet *mass* can be implemented in form of masses of AI-generated images (i.e. masses of information) distributed among a single (or multiple) channel(s). Mass in the information environment corresponds to large quantities of information, which require equally large reserves in analytic resources to make sense out of it. This could be utilized by an actor to selectively overload a channel with misleading, irrelevant or false information. Consequently, the actor's opponent will require some automated process to reduce the mass of (irrelevant) images to be able to identify and process images relevant to them. If no automated process is available, it is likely that the opponent's analytic resources will be strained and overloaded. In practice, this could be applied by an actor to deny one (or multiple) of the opponent's OSINT channels for intelligence collection. Furthermore, it could also be imagined that masses of images are used to bury a particular image instance in it (e.g. one containing leaked sensitive information such as a geolocated air defense asset) and obscure it from the opponent's analysis capabilities or to distract the opponent. As such, masses of AI-generated images can be used to increase the actor's operational security in the information environment. They achieve the tactical tenet of *deception*. In summary, masses of AI-generated images can be used to either deny (or hamper) an opponent's analytic capabilities, or to conceal and distract from activities in the information environment (summarized in Figure 8).

Given an automated production of AI-generated images that are fit for purpose, the tactical tenet *tempo* can be implemented as well. An actor utilizing generative AI can thus be conceptually faster than an opponent relying on traditional image generation. This is relevant for all military activities relying on timely image generation. We know that images grab the audience's attention, they can deceive and manipulate. As outlined in Section 2.2, an actor can therefore use AI-generated images to quickly introduce "entropy" in the opponent's OODA-loop. Consequently, the opponent's OODA-loop will be slowed down, as a result his decision-making process will be slowed down as well, and eventually the pace of combat will be in favor of the actor utilizing AI-generated images. It is important to note that AI-generated images can be combined with other means to introduce "entropy" (e.g. feints, fake radio transmissions, etc.) and that a meaningful combination thereof may result in a greater deterioration of the opponent's OODA-loop. In that sense, AI-generated images may act as a force multiplier that is much faster deployed than a set of images produced by traditional image generation.

Based on the diverse, and easy-to-use interfaces to condition the generation process, there are no bounds set to a generative AI operator's creativity. In combination with the fact that AI-generated images are not grounded in reality, it is easy to generate shocking, disturbing and confusing visual content [34]. The tactical tenet of *surprise* can thus be achieved. In particular, AI-generated images may facilitate the propagation of information for which the opponent is ill-prepared to react to (i.e. no clear Rules of Engagement (RoE), doctrine), leading to surprise on the opponent's side and thus poor decision-making. To illustrate this, we construct a hypothetical scenario: an advanced cyber attack which targets the supply chain of commercially-available satellite images which are used by an opponent. We assume that through this cyber operation the actor can covertly introduce AI-generated satellite images into the image intelligence collection of the opponent. The opponent is ill-prepared to verify the authenticity and quality of the collected satellite images, thus working with information deliberately constructed by the actor to surprise him. Similarly, the tactical tenets *confusion* and *shock* can be achieved as well. For example, confusion can be implemented by affecting the opponents command-and-control capabilities over a prolonged period of time with AI-generated disinformation. In context of our hypothetical scenario, confusion could be caused through a series of bad leadership decisions due to false information extracted from AI-generated satellite images, leading potentially to internal friction, investigations and loss of initiative. According to Friedman, shock is frequently the combination of deception, surprise, and confusion. For our hypothetical scenario, shock may result from actual battlefield losses on the ground caused by a series of bad leadership decisions which were based on information from false AI-generated satellite images (e.g. underestimation of opponent's force presence in a particular area leading to military failure and high casualty rates for that area).

In summary, AI-generated images can be used in masses for deceptive purposes, to influence the pace of combat in the information environment, to cause surprise, sow confusion and shock. In contrast to traditional image generation these tactical tenets can potentially be achieved more efficiently and on a larger scale.

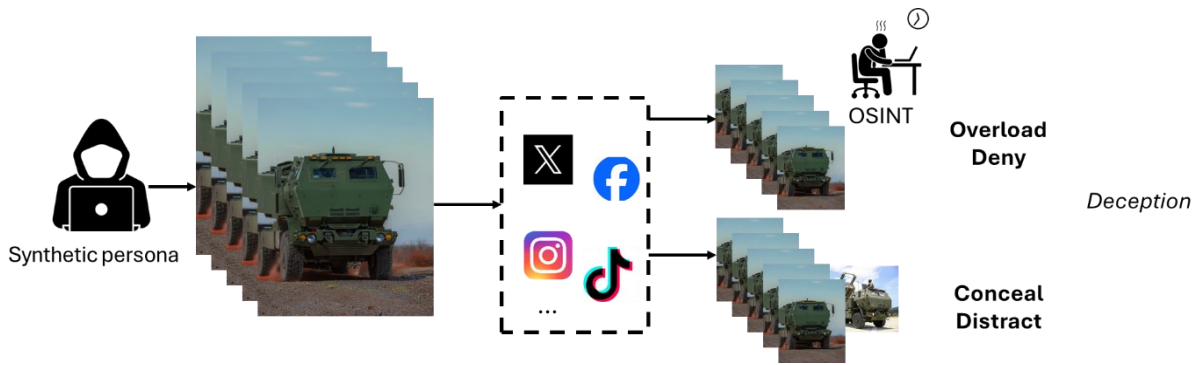

Figure 8. AI-generated images can be used for military deception purposes in the information environment (icons from Flaticon.com, made by Talha Dogar, Leremy, and from wikipedia.org). They can be employed in large masses to either overload the opponent's OSINT capabilities and deny efficient information collection from certain channels, or they can be used to conceal or distract from an unwanted data point (e.g. photograph including sensitive information such as the geolocation of a military asset). Exemplary synthetic and real images of HIMARS rocket launchers for this figure were provided by the authors of [34].

## 4    Opportunities & Threats

In the following, we briefly review a number of opportunities and threats in which AI-generated images may play a role.

### 4.1    Training, Simulation, and Wargaming

Generative AI can be used to produce image data of military assets that in turn can be directly incorporated in existing training activities (e.g. images of military planes used to train air defense operators). Furthermore, AI-generated images could also be incorporated in military exercises and wargaming activities. This could be particularly useful for training activities of military formations that are tasked to engage with the information environment (e.g. analysts of military intelligence branches). Generative AI could be used to generate media content to support simulation of the information environment and potentially also to support drafting the training scenario [35]. For example, NATO strategic communications (StratCom) Centre of Excellence launched a simulation platform for the information environment ("InfoRange") which supposedly incorporates generative AI[15]. The impact of AI-generated images in context of training, simulation, and wargaming will likely be on the tactical level (i.e. generation of images of individual military entities). The role of AI in wargaming in general has been subject to critiques that highlighted the risk of over-trust in AI [36], especially when the generated output is based on an insufficiently credible knowledge base and aims at replacing human decisions [35].

### 4.2    Synthetic Data for Machine Learning Models

AI-generated images can be produced with the goal to augment existing training datasets used for training a variety of ML models to solve different computer vision problems including e.g. image classification [37], image segmentation [38], or depth estimation [39]. Along with the generation of synthetic images using text-to-image models [40], methods based on a combination of techniques including game engines for 3D-models [41] are currently being investigated. There are a multitude of specific applications being researched that are based on synthetic image data and are of particular importance for the armed forces including analysis of satellite images (e.g. detection of airplanes [42] or ships [43]), drone detection [44], vehicle detection [45], image segmentation under adverse weather conditions [46], or support of video surveillance for perimeter security [47]. However, as mentioned already in Section 2.3, the use of synthetic data to train ML models is not without complications. For example, Hennicke et al. [48] showed that naïve transfer learning using synthetic data generated by Stable Diffusion resulted in worse performance than comparable training with real data which persisted even after additional modifications. Similarly, Geng et al. [49] performed a direct comparison of image classification models trained on synthetic data generated by Stable Diffusion, which in turn was trained on LAION-2B dataset, and ML models trained on targeted, retrieved real data from LAION-2B, and found the models trained on synthetic data to consistently underperform. Based on these results, synthetic data should be incorporated in existing training datasets only with great caution and under the supervision of domain experts from the field of machine learning.

---

[15] StratCom | NATO Strategic Communications Centre of Excellence Riga, Latvia [accessed January 2025]

## 4.3 Strategic Communications

Armed forces are continuously exposed in the information environment and perceived by the public through their actions or lack of action. Today's continuum of competition in the information environment requires a constant monitoring of the armed forces' public image, relevant narratives and potentially hostile actions directed against them. To obtain and hold discursive superiority on all relevant issues became a top priority. Strategic communication deals precisely with these challenges. It usually involves also the communication of messages through images. Here AI-generated images may offer additional room for flexibility in creating visual content without the typical resource requirements in traditional image generation (skilled photographer, physical equipment, actors, etc.). Similarly, AI-generated images can be utilized in communication with media and for production of marketing material (e.g. for recruitment efforts). However, this may come at an increased risk of being perceived as "inauthentic" due to the very nature of generative AI. Toff et al. [50] have shown that AI-generated news articles are perceived as less trustworthy than articles written by a human even when articles are evaluated as comparatively accurate and fair. Experiments involving AI-generated public health messaging showed that there is a general disapproval of AI-generated messages compared to messages coming from domain experts (medical doctors) or government bodies (Centers for Disease Control and Prevention) [51]. It is conceivable that other AI-generated government messaging (e.g. by the armed forces) labelled as such is received equally negatively by the target audience. In contrast, a recent study on crisis communication found no significant (though weak negative) effect of source disclosure on message acceptance in a cohort of university students implying different levels of acceptance for different age groups [52]. Given these results, AI-generated content should only be used with great caution in strategic communications.

## 4.4 Deception

Military deception has a long history of using visual deception, e.g. mock airbases and dummy tanks, to deceive and mislead adversaries (e.g. their aerial reconnaissance units). In principle, generative AI would enable actors to generate visual footage of military assets for deception without the need to deploy physical decoys. For example, Mathys et al. (2024) explored the hypothetical scenario in which a threat actor is generating synthetic images of HIMARS rocket launchers. Such images could e.g. be used to generate synthetic visual footage of HIMARS rocket launchers at specific geolocations to mislead an opponent. Furthermore, the opponent could be deceived about the exact number of available HIMARS systems the actor possesses leading to greater uncertainty in decision-making processes. As such, synthetic images of real military assets may serve as a force multiplier. Generative methods which incorporate image edge information (e.g. ControlNet [53]) were identified as particularly potent in such scenarios as they reduce the likelihood of geometrical artefacts or inconsistencies which are typical for the generation of military vehicles (e.g. wrong number of axles, wrong proportionality of launching device versus vehicle, etc.) and would give away the image as being AI-generated. The same conditioning mechanism was also used to condition the generation of satellite images on cartographic information extracted from OpenStreetMap to synthesize satellite images of Scotland [54]. More recently, Liu et al. (2025) proposed a method (Text2Earth [55]) which can generate synthetic satellite images for the whole globe including also the capability to digitally manipulate these images (e.g. selectively adding infrastructure, geographical features, etc.). It is conceivable that generative AI models to synthesize satellite images, particularly when conditioned on real cartographic data, could be used for military deception or cyber influence operations in the near future (e.g. by covertly introducing synthetic data into the satellite image supply chain of the opponent, as mentioned in a previous hypothetical scenario). In summary, military deception executed via the information environment can be enhanced through the use of potentially large quantities of convincingly looking AI-generated images.

## 4.5   Cyber Influence Operations

Actors are increasingly using AI-generated images in cyber influence operations. For example, Microsoft's Threat Analysis Center analyzed the use of AI-generated images in a series of cyber influence operations conducted by actors supposedly aligned with the Chinese Communist Party (CCP)[16]. In these operations, generative AI was used to create both drawings for propagandistic purposes as well as photorealistic content aimed to deceive a target audience. The Stanford internet observatory investigated the activities of supposedly pro-western influence operations[17]. They found GAN-generated images being used to create profile pictures of sockpuppet accounts. Furthermore, AI-generated images could be used in combination with text generated by Large Language Models (LLMs). Recent studies have shown that LLM-generated text can be highly persuasive ( [56], [57], [58], [59]), can exploit the psychological profile of the target audience [60], and is practically impossible to distinguish from human-written text [61]. Research on the persuasiveness, or more generally the psychological impact, of AI-generated images on a target audience is very limited. An already dated study by Moshel et al. [62] showed that while brain activity was detectably different when a person viewed a photograph of a real versus AI-generated face, participants were not able to discriminate real versus AI-generated faces. There is also evidence that AI-generated faces are perceived as more real than human faces, attributed to the hyperrealism of AI-generated images [63]. Interestingly, a more recent study found that participants rated a proportion of real face images as fake with the more attractive faces being rated as more real [64]. This highlights the complexity of how beliefs are shaped in a target audience through image information and underlines our limited understanding of the underlying processes involved in visual perception. Nevertheless, we can assume that AI-generated images, similarly to real images, can capture the audience's attention and may increase the perceived accuracy of the message information. A recent study by Hameleers et al. [65] investigated political deepfake videos for the purpose of delegitimizing a political candidate and found that implausible deepfake videos which were regarded as less credible than plausible videos still had a negative effect on the candidate's perceived legitimacy. Therefore, the deployment of sockpuppet accounts instrumented with generative AI for multi-modal content creation is meanwhile feasible and provides threat actors with a relatively cheap and scalable tool to influence target audiences. In addition, repurposing of existing image material and language discrepancies can be increasingly avoided through the use of generative AI for creation of novel image and text content, which improves the operational security of threat actors conducting cyber influence operations. It is therefore necessary to build up the capacity of cyber defense units to detect, monitor and catalogue the idiosyncrasies of generative AI models left in their output (e.g. visual artifacts produced by text-to-image models). This will enable the implementation of more effective countermeasures for cyber influence operations involving AI-generated images.

## 4.6   Social Engineering

Similarly to techniques for military deception, social engineering profits particularly from the generation of photorealistic synthetic images. The techniques, tactics, and procedures (TTPs) are very similar to what is being employed in cyber influence operations (e.g. reliance on credible sockpuppet accounts) with the main difference of the target being not a social group but an individual person (natural or judiciary). Social engineering leveraged by AI-generated images has been heavily utilized by cyber criminals. Low-quality product scams, romance scams, and insurance fraud are on the rise featuring AI-generated or manipulated images. For example, a series of cases known as "crochet scams"[18] have been targeting persons from the crocheting community, with the sale of non-existent crochet patterns over fake online shops or over third-party platforms (e.g. Etsy). Here, actors utilize generative AI's excellent ability to produce images of items (e.g. crocheting products) which are by eye very hard to distinguish from real products. Furthermore, these actors created a whole ecosystem consisting of fake online shops, fake products and social media groups and fan pages (primarily on Facebook). Interestingly, the reaction of authentic users on social media to AI-generated image content is used as a targeting mechanism to identify supposedly easily manipulatable persons. In other words, if a user reacts (e.g. by leaving a comment) to an AI-generated image, the person is being targeted with a follow-up scam (e.g. romance scam or phishing attack). This case is related to a broader phenomenon referred to as "Facebook engagement spam" which was investigated by DiResta et al. [66] and describes the exploitation of Facebook's recommendation algorithm with attention-seeking AI-generated image content to ramp up engagement in order to launch a multitude of scams and illicit actions. Related to this, are also different criminal activities involving not only AI-generated images but also videos including e.g. the BitGo romance scams involving a mix of AI-generated images and videos to establish trust with the victim and subsequently launch a cryptocurrency scam[19]. Impersonation of high-level executives[20] with the aim to manipulate victims or coerce them into offering sensitive information is an attack vector which also profits from the most recent advances in the generation of photorealistic and personalized synthetic images (e.g. through usage of Dreambooth algorithm [67]). A recent study found that 43% of users connected to fake profiles on social media even when these profiles exhibited obvious artifacts (e.g. inconsistencies of profile picture and biography) and users were informed about how to spot those artifacts. This highlights the potential of social media requests (e.g. on LinkedIn) as attack vector for threat actors, especially when equipped with additional convincing AI-generated images to increase perceived credibility. Consequently, it can be expected that such attack vectors, which are already widely used by cyber criminals, will also be adopted by intelligence services to approach target persons and conduct human intelligence mediated through cyber space.

---

[16] Microsoft Digital Defense Report 2024 [accessed December 2024]
[17] Unheard Voice: Evaluating five years of pro-Western covert influence operations | FSI [accessed September 2024]

## 4.7 Digital Vulnerability

Publicly available, open-source generative text-to-image models (e.g. Stable Diffusion) are particularly suitable for military actors since they are less constrained by their manufacturers (e.g. through filtering of prohibited prompts). Moreover, there is a vibrant online community which continues the development of these models, engages in exchange of tips and tricks, and offers a variety of Graphical User Interfaces (GUI) that enable the use for non-technical operators. However, all this comes at the price of an increased risk of potential attack vectors introduced in either the algorithm, model, or GUI. Recently, the cybercriminal group NullBulge targeted a popular GUI for Stable Diffusion, called ComfyUI, with a malicious extension that when installed exfiltrates the user's credentials[21]. Another known attack vector involves the

---

[18] https://laurarbnsn.substack.com/p/ai-crochet-is-scamming-your-mother [accessed September 2024]
[19] Pig Butchering Scammers Are Using AI Now To Fool Victims – Frank on Fraud [accessed September 2024]
[20] For example, impersonation of the Ukrainian prime minister in a fake video call with Turkish arms producer Baykar (https://www.youtube.com/watch?v=t45HrigEJhk, accessed September 2024).
[21] Hackers Target AI Users With Malicious Stable Diffusion Tool on GitHub to Protest 'Art Theft' [accessed September 2024]

Threats and Opportunities in AI-generated Images for Armed Forcesdownload of malicious model files, typically compressed using Python's pickle module. Hence, the open-source community came up with dedicated scanners that inspect pickled files to avoid integration of malicious models[22]. However, it is likely that threat actors will find novel attack vectors involving the supply chain of generative AI for image generation. Therefore, the integration of generative AI will potentially introduce new digital vulnerabilities for armed forces. In particular, the complex ecosystem and supply chain of open-source tools such as e.g. Stable Diffusion make it very challenging to avoid the introduction of digital vulnerabilities, and their potential benefits should be always weighed against their security risk. Furthermore, best practices from cybersecurity, e.g. compartmentalization, must be applied when implementing generative AI to mitigate potential risks.

### 4.8     Rules of Engagement for Generative AI

A clear set of Rules of Engagement (RoE) for generative AI in armed forces have not been presented yet. The first comprehensive framework on governance of Artificial Intelligence was put forward by the European Union (EU) in their AI Act. However, the AI Act explicitly excludes military AI[23]. Yet, the inherent dual use nature of many AI technologies, including generative AI, make them a prime technology to be adopted by the armed forces. Different initiatives are underway that focus on how generative AI can be used within the military, e.g. in NATO[24] or the US Department of Defense (DoD[25]) but a set of clear RoE have yet to be published. The establishment of RoE has been identified as potential regulatory framework to warrant appropriate usage of generative AI by the armed forces[26]. There is ample room for novel proposals of such RoE and further research into how they should be put into practice for armed forces.

### 4.9     Emergent Threats

With the increasing capability of generative AI to produce photorealistic images, there are multiple emergent threats appearing which can be grouped in two main categories: i) threats aimed at image authenticity, ii) threats facilitating the production and distribution of illegal image content. We will briefly outline the nature of the individual threats in the following.

#### 4.9.1     Increased Confusion on Image Authenticity

It is to be expected that with an increased abundance of synthetic, particularly synthetic photorealistic images, we will see an increased confusion on how image authenticity is defined. There are two recent examples that illustrate this challenge. In the first example, an American political commentator posted a picture on the platform X of the aftermath of the 7th October attack by Hamas on Israel. The picture showed the burnt corpse of a child. It was retweeted by an American influencer showing a supposed screenshot of the image being fed to Optic's "AI or not" detection algorithm available online with the algorithm implying that the image is AI-generated. This stirred up an online debate about the image's authenticity. Furthermore, an alternative version of the image which shows a dog instead of the burnt corpse appeared on the platform 4chan (posted by an anonymous user). Based on this material AlJazeera published the article "A dog, not a child… An American journalist exposes Netanyahu's lies about Hamas killing children". A forensic analysis by the company Adversa proved that in fact the picture showing the dog was digitally altered while the image showing the burnt corpse turned out to be real[27]. The second example that highlights an increased confusion on image authenticity involves a social media post on the platform X by an Israeli journalist[28]. The post shows pictures of the Hamas leadership living a relatively luxurious life when compared with the general population in Gaza. The author decided to upscale the original low-resolution images with a publicly-available upscaling method, which was likely based on a Deep Learning technique. This equipped the images with a synthetic appearance, which backfired at the author when online commentators were claiming that he was sharing AI-generated fake images.

---

[22] GitHub - diStyApps/Stable-Diffusion-Pickle-Scanner-GUI: Pickle Scanner GUI
[23] Governing Military AI Amid a Geopolitical Minefield | Carnegie Endowment for International Peace [accessed November 2024]
[24] Military Applications of Generative AI [accessed November 2024]
[25] Chief Digital and Artificial Intelligence Office > Initiatives > AI Rapid Capabilities Cell [accessed November 2024]
[26] Rules of Engagement as a Regulatory Framework for Military Artificial Intelligence - Lieber Institute West Point [accessed November 2024]

Based on these two examples, it is evident that confusion on image authenticity will only increase in the future. In particular, incorrect predictions by publicly-available AI image detection methods can be used to discredit real photographs as fake. This adds to the already problematic phenomenon of the "liar's dividend": the repeated strategic use of false and/or synthetic media may increase a general scepticism towards objective truth. Consequently, threat actors can more easily dismiss authentic images as fake and avoid accountability (e.g. for atrocities committed during a war) [34].

### 4.9.2 AI-generation of Illegal Images

Generative AI models can also be used by threat actors to generate illegal image content including images containing graphical violence, child sexual abuse material (CSAM), or non-consensual nudity. Text-to-image models are usually trained on very large image-text datasets scrapped from the internet. Given the number of samples being in the billions, large extents of these datasets were never manually reviewed (e.g. for the presence of illegal image material). For example, the frequently employed LAION-7B which e.g. was used to train Stable Diffusion has been shown to contain illegal CSAM content [68] [69]. Consequently, the models trained on such data have the potential to reproduce equally illegal image content. Text-to-image models which are hosted by commercial vendors, e.g. Microsoft's Bing Image Creator, usually operate on the level of text prompts to filter out any generation of illegal or harmful image content. Threat actors may try to evade such filtering through the means of prompt engineering[29]. Therefore, attempts to prevent the generation of illegal or harmful image content in generative AI by design (i.e. algorithmically) is an active area of research but it has yet to be shown that an effective solution exists for filtering out illegal or harmful concepts (e.g. CSAM).

In context of armed conflict, the generation of graphical violence could be used by actors in cyber influence operations to undermine the moral cohesion of an opponent (e.g. showing fake images of killed and wounded enemy troops). Furthermore, fake non-consensual nudity may play a role for offensively acting intelligence services in facilitating coercion of target persons. Hence, raising awareness on the capabilities of generative AI for image creation and providing

---

[27] More details can be found here: https://adversa.ai/blog/aljazeera-fake-news-investigation-burned-babies-and-an-ai-generated-dog-images/ [accessed October 2024].
[28] More details can be found here: https://www.forbes.com/sites/mattnovak/2023/10/21/viral-photos-of-hamas-leaders-accused-of-being-ai-fakes-actually-just-poorly-upscaled/ [accessed October 2024].
[29] The Folly of DALL-E: How 4chan is Abusing Bing's New Image Model - bellingcat [accessed October 2024]

education on how to differentiate synthetic photographs from real photographs are essential to increase self-protection of troops in cyberspace and the information environment.

### 4.9.3 Synthetic History

Generative AI shows an ever-increasing capability to generate photorealistic image content. Furthermore, recent models offer the generation of photorealistic images which exhibit particular visual characteristics that are representative for a historic period (e.g. black-white photographs, grainy mobile photographs of the late 2000's, etc.). This was recognized by an online community interested in artistically exploring alternative or synthetic histories (e.g. a history in which Nazi Germany won the second world war[30]), which utilizes text-to-image models to create image (or video) content of entirely fabricated synthetic historical events (e.g. 2001 Great Cascadia Earthquake & Tsunami[31]).

The capability to equip entirely fabricated synthetic historical events with image (and video) content could be weaponized for the strategic manipulation of public perception over extended periods of time [70]. Autocratic regimes could use this capability to e.g. rewrite their own history, lend support to conspiracy theories, and ultimately generate a credible narrative to prepare their population for war against an enemy.

---

[30] Synthetic videos can be found here: Epentibi - YouTube. [accessed November 2024]
[31] AI Creates Photo Evidence Of 2001 Earthquake That Never Happened [accessed November 2024]

## 5   Future Technological Evolution

The pace of research and development in the field of AI is rapid and steadily increasing with 149 foundation models being released in 2023, which is more than double the number of releases in 2022 [71]. New models may feature new capabilities, which in turn may lead to new tactical opportunities for armed forces. We focus on highlighting two main current limitations of generative AI for image synthesis as a starting point to discuss the potential future technological evolution. Any development that may overcome such limitations will have a considerable impact on the operative use of generative AI.

It has been shown that a phenomenon called "concept fusion" frequently occurs for state-of-the-art text-to-image models [34]. It describes the duplication of unique entities (e.g. persons) or separate elements (e.g. different clothing) into a composite. This highlights the fact that there is no precise control over how concepts defined in the text prompt are spatially mapped to regions in the generated image. The authors of [34] observed (on a limited sample) that concept fusion occurs more frequently for fine-tuned models, which may indicate a relationship with overfitting. Novel developments which can mitigate concept fusion or enable a more fine-grained control over how concepts in the text prompt are mapped to the generated image will have a considerable impact on the quality and reliability of image synthesis. Recently, the method ObjectDiffusion has been proposed [72], which combines the text-to-image model of Stable Diffusion with a condition mechanism that maps objects in the text prompt (e.g. car or dog) to user-defined bounding boxes in the image. Hence, we expect to see more research like this in the future leading to an increased ability to condition and control the image generation process, which will eventually mitigate the issue of concept fusion entirely. As a consequence, there will be less synthesized images with artifacts stemming from concept fusion, which will decrease the number of required generations to identify a suitable synthetic image that fits the pre-defined criteria of the human operators (i.e. an increased efficiency in generation of task-specific synthetic images is to be expected).

As outlined in Section 3, the assessment of fit for purpose of the generated candidate images has still to be performed manually by a human operator. This is a time-consuming and cumbersome process. While the evaluation of presence of objects in an image can be automated through the means of an object detection algorithm, it is not yet clear how emotions that may be induced by a synthetic image can be evaluated automatically. The research field concerned with the identification and processing of human emotions using AI is affective computing. Recent advances in this field led to Large Language Models (LLMs) being capable of recognizing human emotions in text [73], which is a basic sign of cognitive empathy for such models. Further studies have shown capability of LLMs for recognition of suitable behavior in social situations [74] (e.g. how to handle a conflict at work), and the ability to recognize emotions from the eyes' region of facial images across different ethnicities [75]. The extension of these results to AI-generated images with arbitrary content will benefit automatic evaluation of fit for purpose for synthetic images. Effective combination of LLMs for automatic evaluation of synthetic images produced by text-to-image models appears currently as the most attainable option. However, LLMs struggle with more advanced general-purpose tasks beyond affective computing that may be useful for such image evaluation tasks, e.g. inductive reasoning remains poor in LLMs [76]. If automatic evaluation of synthetic images becomes feasible in the near future, it will have a considerable impact on tactical concepts. In particular, it would remove the requirement of a human operator reviewing candidate images, reducing the time from image generation to deployment of synthetic images. This would greatly benefit offensive manoeuvres in context of deception, cyber influence operations, and social engineering. The presented tactical concepts in Figure 8, which build on masses of synthetic images, will be much easier to implement in the information environment. Automated synthetic image generation, evaluation, and deployment in context of cyber influence operations will become particularly potent when combined with methodologies used for algorithmic cognitive warfare including fine-grained target audience identification [77], and prediction of public opinion [78].

In summary, we expect research and development in generative AI to further increase the level of quality of the synthesized images and the degree of automation of the image generation process and subsequent evaluation tasks. As a consequence, armed forces will potentially be faced with a higher intensity of offensive manoeuvres, both in terms of volume and speed, using deceptive synthetic images in the information environment.

# 6   Conclusion and Recommendations

AI-generated images are part of a series of technological breakthroughs that shaped today's information environment, making it particularly vulnerable to manipulative, deceitful and subversive actions. They can be employed in both defensive and offensive applications leading to several opportunities and threats for armed forces. While AI-generated images adhere to the two basic military functions (sensor output, effector on cognition) similar to conventional (e.g. photographic or drawn) images, they offer new, distinct conceptual advantages: i) relatively simple to use mechanisms for conditioning the generation process, and ii) scalability. The former includes different interfaces, such as e.g. textual prompts or image information that can be used to influence the generation process, offering great flexibility to tailor the output image to intended purpose and other predefined criteria (e.g. image style, emotionality, etc.). The latter is based on the fact that generative AI is implemented programmatically and in principle can be run automatically to generate masses of synthetic images. The only remaining bottleneck, which at the moment can only be partially automated, lies in the qualitative evaluation of the generated images (i.e. verifying if the image fits the intended purpose and predefined criteria). AI-generated images can be produced more quickly and with greater flexibility when compared to traditional image generation (e.g. photography and image editing). It can be expected that large masses of AI-generated images will be used to either deny information channels to an opponent or conceal sensitive information. AI-generated images may influence the pace of combat through increasing entropy in the opponent's OODA-loop, cause surprise, sow confusion and shock. These tactical tenets can also be achieved using images produced with traditional image generation methods, but generative AI can achieve it more efficiently and on a larger scale. Synthetic images created with generative AI may be used to enhance already existing training datasets for a diverse range of machine learning methods (e.g. for object detection, image segmentation, etc.). This has the potential to improve the performance of any sensor, which is built on the combined use of image data and machine learning analysis (e.g. automatic analysis of satellite images). A much greater impact will, however, be seen for offensive applications aimed at manipulating, deceiving or subverting target audiences. Offensive applications in the areas of deception, cyber influence operations, and social engineering will particularly profit from AI-generated images. Actors involved in such activities do not require the repurposing of old image material anymore, may tailor their images to very specific purposes (e.g. targeting single individuals), and may exploit the vulnerability of recommender systems on social media to push AI-generated image content. However, they may also face considerable challenges due to the inherent trial-and-error process of creating AI-generated images, the difficulty to achieve convincing photorealism, and the required experience and knowledge to employ generative AI tools for synthetic image generation. Furthermore, AI-generated images have not led to entirely new threats so far but enhanced already existing ones (e.g. low-quality product scams) with one exception: cybercriminals have started to use the reaction of users to AI-generated images on social media as a targeting mechanism to identify particularly vulnerable persons for follow-up attacks (e.g. romance scam, phishing attacks). It can be expected that this tactic will be coopted by state-sponsored actors involved in cyber influence operations as well. We recommend answering this increased offensive potential with several distinct countermeasures to increase self-protection of the armed forces in the information environment:

- Continuous monitoring of emerging generative AI models, capabilities, and private entities. For the latter, private entities offering "influence-for-hire" services involving synthetic media generation are of particular concern. By knowing the technological state-of-the-art and its usage in industry, armed forces have the necessary situational awareness to adequately respond to emerging threats.

- Based on the best performing generative AI methodology and models to generate synthetic images, armed forces should actively engage in red teaming exercises relevant for their area of responsibility. Such exercises may be helpful in obtaining estimates on how attainable particular attack vectors are for threat actors, how well current countermeasures are working, and improve interoperability among different units exposed to the information environment.
- Armed forces must be capable to distinguish AI-generated from real photographs. Hence, manual approaches based on the identification of visual artifacts as well as automatic detection with machine learning have to be implemented. For analysis of high-stakes image data, a combined approach of automatic processing and manual expertise by analysts may be the most effective measure. In general, automatic detection of AI-generated images will always result in some images being misclassified and as such is insufficient to handle this classification task autonomously.
- Units concerned with cyber defense should actively explore and prepare guidelines and countermeasures for cyber influence operations involving large quantities of synthetic images.
- Intelligence analysts need to be continuously trained in the forensic analysis of AI-generated images. In particular, manual verification of image authenticity based on the analysis of visual artifacts must be part of the training program.
- Troops should be educated on the current threat landscape involving synthetic media content, which includes AI-generated images. Training to increase general media competency, adherence to information security guidelines and promotion of a mindset geared towards data protection are needed.

In summary, AI-generated images function primarily as a force multiplier for several cognitive threats in the information environment. Armed forces are required to harden their cyber defense against these threats and increase self-protection of troops through adequate education and training.

Threats and Opportunities in AI-generated Images for Armed Forces[41] S. Hinterstoisser, O. Pauly, H. Heibel, M. Marek und M. Bokeloh, „An Annotation Saved is an Annotation Earned: Using Fully Synthetic Training for Object Instance Detection," February 2019.

[42] J. Downes, W. Gleave und D. Nakada, „RarePlanes Soar Higher: Self-Supervised Pretraining for Resource Constrained and Synthetic Datasets," in *Proceedings of the IEEE/CVF Winter Conference on Applications of Computer Vision (WACV) Workshops*, 2023.

[43] C. M. Ward, J. Harguess und C. Hilton, „Ship Classification from Overhead Imagery using Synthetic Data and Domain Adaptation," in *OCEANS 2018 MTS/IEEE Charleston*, 2018.

[44] T. R. Lenhard, A. Weinmann, K. Franke und T. Koch, „SynDroneVision: A Synthetic Dataset for Image-Based Drone Detection," November 2024.

[45] T. A. Eker, E. P. Fokkinga, F. G. Heslinga und K. Schutte, „Balancing 3D-model fidelity for training a vehicle detector on simulated data," in *Artificial Intelligence for Security and Defence Applications II*, 2024.

[46] B. Gella, H. Zhang, R. Upadhyay, T. Chang, M. Waliman, Y. Ba, A. Wong und A. Kadambi, „WeatherProof: A Paired-Dataset Approach to Semantic Segmentation in Adverse Weather," December 2023.

[47] J. R. Waite, J. Feng, R. Tavassoli, L. Harris, S. Y. Tan, S. Chakraborty und S. Sarkar, „Active shooter detection and robust tracking utilizing supplemental synthetic data," September 2023.

[48] L. Hennicke, C. M. Adriano, H. Giese, J. M. Koehler und L. Schott, „Mind the Gap Between Synthetic and Real: Utilizing Transfer Learning to Probe the Boundaries of Stable Diffusion Generated Data," May 2024.

[49] S. Geng, C.-Y. Hsieh, V. Ramanujan, M. Wallingford, C.-L. Li, P. W. Koh und R. Krishna, „The Unmet Promise of Synthetic Training Images: Using Retrieved Real Images Performs Better," June 2024.

[50] B. Toff und F. M. Simon, „"Or They Could Just Not Use It?": The Dilemma of AI Disclosure for Audience Trust in News," *The International Journal of Press/Politics,* December 2024.

[51] E. Karinshak, S. X. Liu, J. S. Park und J. T. Hancock, „Working With AI to Persuade: Examining a Large Language Model's Ability to Generate Pro-Vaccination Messages," *Proceedings of the ACM on Human-Computer Interaction,* Bd. 7, p. 1–29, April 2023.

[52] E. C. Ray, P. F. Merle und K. Lane, „Generating Credibility in Crisis: Will an AI-Scripted Response Be Accepted?," *International Journal of Strategic Communication,* p. 1–18, December 2024.
31

Threats and Opportunities in AI-generated Images for Armed Forces